\documentstyle[jeep,graphics,equations,twocolumn,doublespace]{article}

\begin{document}           

\newpage

\begin{center}

{\Large {\bf Rapid parapatric speciation on holey adaptive landscapes}}\\ 

\vspace{.25in}

{\large Sergey Gavrilets$^{1,2,*}$, Hai Li$^3$, Michael D. Vose$^3$} \\  

\end{center}

Departments of $^1$Ecology and Evolutionary Biology, 
$^2$Mathematics, and $^3$Computer Science,
University of Tennessee, Knoxville, TN 37996, USA.
 
\vspace{.25in}
$^*$To whom correspondence should be addressed. Email: gavrila@tiem.utk.edu\\

Brief title:  Rapid Speciation\\


\vspace{.25in}

\section{ SUMMARY}

A classical view of speciation is that reproductive isolation arises as
a by-product of genetic divergence. 
Here, individual-based simulations are used to evaluate whether the 
mechanisms implied by this view may result in rapid speciation
if the only source of genetic divergence are mutation and random genetic drift.
Distinctive features of the simulations are the consideration 
of the complete process of speciation (from initiation until completion), and
of a large number of loci, which was only 
one order of magnitude smaller than that of bacteria.
It is demonstrated that rapid speciation on the time scale of hundreds of generations is plausible without the need for extreme founder 
events, complete geographic isolation, the existence of distinct adaptive
peaks or selection for local adaptation.
The plausibility of speciation is enhanced by population subdivision.
Simultaneous emergence of more than two new species from a subdivided 
population is highly probable.  
Numerical examples relevant to the theory of centrifugal speciation and to 
the conjectures about the fate of ``ring species'' and ``sexual continuums''
are presented.

\section{ 1. INTRODUCTION}

In spite of significant advances in both theoretical and experimental
studies of evolution, understanding speciation still remains a major challenge (Mayr 1982; Coyne 1992).
In recent years a view of speciation that can be traced to the  classical 
Dobzhansky model (Dobzhansky 1937; Muller 1942) has become a point of renewed 
interest (Nei {\em et al.}  1983; Wagner {\em et al.}  1994; Orr 1995; Orr \& Orr 1996;
Gavrilets 1997a,b; Gavrilets \& Hastings 1996; Gavrilets \& Gravner 1997). 
Using the metaphor of ``adaptive landscapes'' (Wright 1932), populations 
diverge genetically along the ``ridges'' of highly-fit genotypes 
and become
reproductively isolated species when they come to be on opposite sides of 
a ``hole'' in a ``holey'' adaptive landscape (Gavrilets 1997a; Gavrilets
\& Gravner 1997).

Current renewal of interest in Dobzhansky-type models is coming 
from different sources. First, there is growing experimental support for the
genetic architecture implied (reviewed in Orr 1995; Orr \& Orr 1996; Gavrilets 1997a,b).
On the theoretical side, there is a growing realization of serious problems 
with Sewall Wright's metaphor of ``rugged'' adaptive 
landscapes (Whitlock {\em et al.}  1995; Gavrilets 1997a)
and shifting balance theory (Gavrilets 1996; Coyne {\em et al.}  1997).
Both have for a long time have been a focus of theoretical studies of evolution.
On the other hand, it recently has been demonstrated 
that the existence of ``ridges'' of highly-fit genotypes, which is postulated
in Dobzhansky-type models, is actually a general property of
multidimensional adaptive landscapes (Gavrilets \& Gravner 1997; Reidys
{\em et al.}  1997).

Earlier theoretical studies of speciation as a consequence of genetic 
divergence along ``ridges'' in the adaptive landscape have shown it to be
a distinct possibility. These studies involved various approximations such 
as small number of loci
(Nei {\em et al.}  1983; Wagner {\em et al.}  1994; Gavrilets \& Hastings
1996), linkage equilibrium (Nei {\em et al.}  1983), infinitely 
large number of highly mutable unlinked loci and preferential mating 
(Higgs \& Derrida 1992;
Manzo \& Peliti 1994), or absence of within-population variability during
divergence (Orr \& Orr 1996). Here we extend the previous work and report
results of individual-based 
simulations that did not involve any approximations typical in studying 
multilocus systems. Using a linear algorithm for generating 
random numbers with a given distribution (Vose 1991) has allowed us to 
model hundreds of linked loci potentially influencing fitness. 
The number of considered loci, which is only one order 
of magnitude smaller than that of bacteria (Blattner {\em et al.}  1997), 
is a significant distinctive feature of our simulations. An analytic theory 
for the dynamic patterns observed will be published elsewhere 
(Gavrilets unpublished).

\section{ 2. MODEL}

We model finite subdivided populations of sexual haploid individuals different
with respect to $L$ diallelic loci. Evolutionary factors included were 
mutation, recombination, migration, 
genetic drift and selection. To decrease the number of parameters
we used some symmetry assumptions. Mutation rates were equal for forward 
and backward mutations and across loci. Recombination rates between
adjacent loci on the same chromosome were equal and recombination events 
took place independently of each other. The population was subdivided into
subpopulations of equal size.

Without any loss of generality each individual can be represented as a 
sequence of 0's and 1's. Let $l^{\alpha}=(l_1^{\alpha},...,l_L^{\alpha})$ 
where $l_i^{\alpha}$ is equal to 0 or 1, be such a sequence for an individual
$\alpha$. We will be interested in 
the levels of genetic variability within subpopulations and genetic divergence
between subpopulations. Both can be characterized in terms of genetic distance
$d$ defined as the number of loci at which two individuals are different. 
More formally, the genetic distance $d^{\alpha \beta}$ between individuals
$\alpha$ and $\beta$ is
   \begin{equation} \label{distance}
          d^{\alpha \beta} = \sum_{i=1}^L (l_i^{\alpha}-l_i^{\beta})^2.
   \end{equation} 
Genetic distance $d$ is the standard Hamming distance. To characterize
the levels of genetic variability within subpopulations and genetic divergence
between subpopulations we used the average genetic distances (AGD) between individuals from the same subpopulation and from different subpopulations, 
respectively. To reflect the idea that reproductive isolation arises simultaneously with genetic divergence we posit that an encounter of 
two individuals can result in mating and viable and fecund offspring only 
if the individuals are different in no more than $K$ loci (cf., Higgs \&
Derrida 1992).
Otherwise the individuals do not mate (premating reproductive isolation)
or their offspring is inviable or sterile (postmating
reproductive isolation). More formally, we assign
``fitness'' $w$ to each pair of individuals depending on the genetic
distance $d$ between them
    \begin{equation} \label{thres}
         w(d) = \left\{ \begin{array}{cc}
                        1 & \mbox{for $d\leq K$},\\
                        0 & \mbox{for $d>K$}.
                       \end{array}
                \right.
    \end{equation}
In this formulation, any two genotypes different in more
than $K$ loci can be considered as sitting on opposite sides of a hole
in a holey adaptive landscape. At the same time, a population can evolve
to any reproductively isolated state by a chain of single locus
substitutions. The neutral case (no reproductive isolation) corresponds 
to $K$ equal to the number of loci.
In our simulations individuals migrated before mating and pairing of 
individuals occurred randomly. There were $384$ on a single
chromosome, the rate of recombination between adjacent loci was 
$0.005$, and the mutation rate per locus was $10^{-4}$.
The $K$ values used are 10, 20 and 30 which 
are within the range of estimates of the minimum number of genes 
involved in reproductive isolation (Singh 1990; Coyne and Orr 1998).

Our simulations started with all $N$ individuals identical. During the first
1000 generations there were no restrictions on  migration  
between subpopulations and the whole population evolved as a single 
randomly mating unit. One thousand generations was sufficient for the 
population to reach a state of stochastic equilibrium between factors
affecting its dynamics. This was apparent from the behavior of AGD 
(see figures below) as well as from the behavior of the number of
segregating loci and the number of different genotypes (data not presented).
Starting with generation 1000,  restrictions on migration 
were introduced. We used both the island model (Wright 1931) and the
one-dimensional stepping-stone model (Kimura \& Weiss 1964). In the former case, with probability 
$m$ an individual migrated to another subpopulation (with all subpopulations 
equally accessible). In the later case, the subpopulations were arranged along
a line with migration only between neighboring subpopulations (with rate
$m/2$). The simulations
were continued until generation 3000. We have chosen this relatively short
time span for two reasons. First, here we are  interested in the
 plausibility of {\em rapid} speciation. Second, only during 
relatively short time intervals
may one assume the constancy of abiotic and biotic environment implied in
our model.

\section{ 3. NUMERICAL RESULTS}

Extensive simulations have revealed the 
existence of three different dynamical regimes. In the first regime, the AGD
both between and within subpopulations reach (stochastic) 
equilibrium values below $K$ (Fig.1a). Between-subpopulations matings take
place and are 
fecund. No further genetic divergence and speciation take place (at least 
until generation 3000). 
In the second regime, the AGD between subpopulations steadily increases 
after introduction of restrictions on migration (Fig.1b). 
Subpopulations diverge genetically accumulating different mutations. 
After some time, three effects take place simultaneously:
(1) AGD between subpopulations significantly exceed AGD within them, (2) encounters between individuals from different 
subpopulations do not result in viable and fertile offspring, (3) evolutionary
changes in a subpopulation do not affect other subpopulation.
Thus, subpopulations form separate genotypic clusters in genotype space,
become reproductively isolated and undertake changes as evolutionary 
independent units. These dynamics are interpreted as speciation according to 
any of the species concept common in the literature
(e.g., Mallet 1995; Claridge et al. 1997). Here, speciation has 
occurred in the presence of gene flow and, thus, is parapatric.
The third regime is a combination of the first two. After
some transient time during which the AGD both within 
and between subpopulations have appeared to reach 
a new stochastic equilibrium, the AGD between subpopulations starts suddenly increasing (Fig.1c). 
The system undergoes a stochastic transition resulting in speciation 
and further accumulation of genetic differences between subpopulations. 
After speciation has taken place, the genetic variability within
(sub)populations decreases. [Note that instead of the three different
dynamical regimes one may choose to think of a single regime with a 
varying (from zero to infinity) time until the initiation of divergence.]

Dynamic behavior depends on the intensity of different
factors operating in the system. As expected, reduced migration and increased mutation both increase the plausibility of speciation (data not presented).
Population size and the strength of selection play a double role.
With strong selection (small $K$), increased population size
delays speciation (compare Fig.2a and b). 
With weak selection (large $K$), decreasing population size delays
speciation (compare Fig.2c and d).
With smaller population sizes, decreasing the strength of
selection delays speciation (compare Fig.2a and c). 
With larger population size, increasing the strength of selection seems to
delay speciation (compare Fig.2b and d).

These effects can be explained in the following way. A necessary 
condition 
for genetic divergence and speciation is fixation of alternative alleles 
in different subpopulations. After
the (sub)population becomes polymorphic at $K$ loci, new mutations
are selected against because individuals carrying them have a reduced 
probability of fecund mating. Genetic drift operating in small populations
overcomes the effect of selection and allows for genetic divergence. 
Genetic divergence of very large randomly mating subpopulations connected 
by migration will 
not occur. On the other hand, if (sub)populations are too small, they lack genetic variability necessary to initiate divergence. Genetic divergence of 
very small randomly mating subpopulations connected by migration will not 
occur either. 
The dynamics of genetic divergence 
are not neutral. During the phases of stochastic equilibrium the AGD 
both within and between subpopulations are well below neutral 
expectations (which are given in Watterson 1975; 
Li 1976; Slatkin 1987; Strobeck 1987). 
After speciation has taken place, the rate of between-population divergence
is smaller than the neutral expectation (which is twice the expected number 
of mutations per generation, $v$).

Increased population subdivision promotes speciation.
For example, in a population with $N=800$ subdivided into two subpopulations
no speciation has been observed (Fig.3a).  However, subdividing 
into four ``islands'' immediately initiates divergence 
and emergence of 4 species (Fig.3b). The migration rate in these 
simulations was small ($m=0.01$). Speciation can
also occur for higher migration rates with appropriate spatial subdivision. Fig.3c shows results for a single run of a stepping-stone model with 
8 subpopulations and $m=0.06$ (that is there are 6 migrants per subpopulation per generation). Here, after some transient time the population splits into 
2 species. Splits have occurred in 70 runs (out of 92 performed),
in 18 of which the population was split into three species.
Figure 4a shows that the two peripheral populations have reduced probability
of separating from the whole system. Figure 4b suggests that this is so
because peripheral populations on average have reduced genetic variability
and are more similar to their neighbors relative to other populations.
Fig.4c shows that, as expected, genetic distance increases with the geographic
distance. The AGD between the first subpopulation and subpopulations four
through eight is above $K$ meaning these pairs of subpopulations are reproductively isolated. However, they evolve together and have a common 
evolutionary trajectory.

\section{4. DISCUSSION}

Here, individual-based simulations were used to evaluate the plausibility
of rapid speciation as a result of genetic divergence along the ridges of
a holey adaptive landscape. Distinctive features of the simulations are the 
consideration of the complete process of speciation (from initiation until
completion), and of a large number of loci, which was only 
one order of magnitude smaller than that of bacteria. Our numerical results 
have implications for several important evolutionary questions
that have been extensively discussed in the literature but have not been
approached using explicit mathematical models.  

{\em 1. Rapid speciation is plausible without the need for extreme founder 
events, complete geographic isolation, the existence of distinct adaptive
peaks or selection for local adaptation.}
In our simulations, speciation was observed when the migration rate was on 
the order of several individuals per subpopulation per generation.
The time scale in the simulations was short, meaning that 
restrictions on migration should not be long-lasting. Several hundreds 
generations might be sufficient for a significant divergence and evolution
of reproductive isolation (Fig.1,2,3). Restoring migration after that to 
higher levels 
will not return the system back to its initial state of free gene exchange
between subpopulations. The rates of speciation observed here are much higher 
than in models of speciation via transitions between adaptive peaks 
across valleys of maladaptation (e.g. Barton and Rouhani 1993) and in
earlier models of holey adaptive landscapes that considered only few loci 
(Nei {\em et al.}  1983; Wagner {\em et al.}  1994; Gavrilets \& Hasting 1996). 
In our model, random genetic drift and mutations were the only source of 
genetic divergence and speciation.
Here we tacitly assumed that the loci controlling reproductive
isolation are different from those responsible for local adaptation. 
If this were not the case or if there were close linkage between the two types 
of loci, selection for local adaptation might accelerate genetic divergence
and speciation (e.g. Schluter 1996).

{\em 2. The plausibility of speciation is enhanced by population subdivision}.
In our model, increasing the number of subpopulations makes speciation more 
plausible (Fig.3).
The number of subpopulations can be interpreted as a measure of the
geographic range size of the population. Thus, our results show that 
populations with larger geographic range sizes have a greater probability 
of speciation (cf, Rosenzweig 1995). Our conclusion about the effect of 
population subdivision on the probability of speciation in Dobzhansky-type models differs from that of
Orr and Orr (1996). They argued that the degree of population subdivision
has no effect on the rate of speciation if speciation 
is caused by mutation and random drift. The problematic assumption of 
their analysis is
that fixation of new mutations is a neutral process. However, the existence
of holes in the adaptive landscape makes the process of substitution  
non-neutral and new mutations are selected against when rare (see above). 
Such mutations are fixed more easily in smaller subpopulations (Fig.2).

{\em 3. Simultaneous emergence of several new species from a subdivided
population is highly probable}. Hoelzer and Melnick (1994) have emphasized 
that this possibility should be incorporated more explicitly in 
the contemporary methods for reconstructing phylogenies.
Recently several cases of rapid speciation have been described 
(Schluter \& McPhail 1992; McCune 1996; Johnson {\em et al.}  1996). 
The most spectacular one is probably the origin of hundreds of species of 
Lake Victoria cichlids in 12,000 years (Johnson {\em et al.}  1996). 
Our simulations have provided examples of two, three and four new 
species emerging from a subdivided population during a short time interval. Numerical examples
with hundreds of new species can be constructed by increasing the
number of subpopulations. Overall our simulation model provides a justification
for a verbal ``micro-allopatric'' model of speciation suggested for cichlid fishes (e.g., Reinthal and Meyer 1997).

{\em 4. Peripatric speciation vs. centrifugal speciation}. 
Under appropriate conditions, a subdivided population will 
split into two or more distinct species. An interesting question concerns
the likelihood of the split at different points along the geographic range 
of the species (e.g. Gaston 1998). Mayr's (1952) theory of peripatric 
speciation argues that peripheral populations are more likely to split off. 
Surprisingly, our simulations of a one-dimensional stepping-stone model
show that the breaks resulting in the splitting of peripheral populations 
have the lowest probability (Fig.4a). A reason for this is that peripheral
populations have lower genetic variability and are more similar to their
neighbors than the ``inner'' subpopulations (Fig.4b). That genetic changes
resulting in speciation are more probable in the middle of the species 
range was argued by Brown (1957) in his theory of centrifugal speciation.
It would be interesting to see if the splitting pattern that we have observed
stays the same in more complex models of spatial arrangement of subpopulations.

{\em 5. ``Ring species'' and the instability of the ``sexual continuum.''}
About one forth of the runs of the stepping-stone model with
8 subpopulations did not result in speciation within the time interval 
studied. In these runs, the neighboring subpopulations were not
reproductively isolated. At the same time, the AGD between peripheral 
populations significantly exceeded the threshold value $K$ (Fig.4c) meaning 
that if these populations somehow came into a direct contact they would be 
reproductively isolated. The situation observed here is analogous to that 
in ``ring species'' which are traditionally considered as examples of incipient, but incomplete, speciation (e.g., Mayr 1942, 1952; Wake 1997). 
The population as a whole is characterized by the existence of reproductively
isolated groups that exchange genes through a series of intermediate genotypes.
Maynard Smith and Sz\"{a}thm\'{a}ry (1995, pp. 163-167), who called such a population state ``sexual continuum'', conjectured that ``sexual continuum'' 
is unstable and would break up into distinct ``species.''
Noest (1997) has proposed a model in which a uniform ``sexual continuum''
was unstable to a range of perturbation. His analysis did not address
whether the system will evolve to a non-uniform ``sexual continuum'' or to 
a discontinuous state.
Although the ``sexual continuum'' observed in the runs described by Fig.4c
appears to be stable, continuing simulations for a longer period of time 
would almost definitely result in speciation in all runs. 
This provides a justification for Maynard Smith and Sz\"{a}thm\'{a}ry's conjecture. 
Developing an analytic theory for predicting the persistence time of a 
``sexual continuum'' and ``ring species'' is an important direction for
future work.\\ 


{\small We are grateful to Mark Camara, Mitch Cruzan, John Gittleman,
Massimo Pigliucci and anonymous reviewers for helpful comments and suggestions. S.G. was partially supported by
a Professional Development Award and EPPE fund (University of Tennessee, Knoxville), by 
grants from Universit\'{e} P. et M. Curie and \'{E}cole Normale Sup\'{e}rieur,
Paris, and by National Institutes of Health grant GM56693.}\\


\section{ REFERENCES}

 Barton, N.H. \& Rouhani, S. 1993. Adaptation and the
``shifting balance.'' {\em Genet. Res.} {\bf 61}, 57-74.

  Blattner, F.R. {\em et al.} 1997. The complete genome sequence
of {\em Escherichia coli} K-12. {\em Science} {\bf 277}, 1453-1462.

  Brown, J.L. Jr. 1957. Centrifugal speciation. {\em Quart. Rev. Biol.}
{\bf 32}, 247-277.

  Coyne, J.A. 1992. Genetics and speciation.{\em Nature} {\bf 355}, 511-515. 

  Coyne, J.A., Barton, N.H. \& Turelli, M. 1997.
A critique of Sewall Wright's shifting balance theory of evolution.
{\em Evolution} {\bf 51}, 643-671.

  Coyne, J.A. \& H.A. Orr. 1998. The evolutionary genetics of
speciation. {\em Phil. Trans. R. Soc. Lond. B} {\bf 353}, 287-305.

  Dobzhansky, Th.G. 1937. {\em Genetics and the Origin of Species}. 
New York: Columbia University Press. 

  Gaston, K.J. 1998. Species-range size distributions: products of
speciation, extinction and transformation. {\em Phil. Trans. R. Soc. Lond.} 
B {\bf 353}, 219-230.

  Gavrilets, S. 1996. On phase three of the shifting-balance 
theory. {\em Evolution} {\bf 50}, 1034-1041.

  Gavrilets, S. 1997a. Evolution and speciation on holey adaptive
landscapes. {\em Trends Ecol. Evol.} {\bf 12}, 307-312.

  Gavrilets, S. 1997b. Hybrid zones under Dobzhansky-type
epistatic selection. {\em Evolution} {\bf 51}, 1027-1035.

  Gavrilets, S. \& Gravner, J.  1997.  Percolation 
on the fitness hypercube and the evolution of reproductive isolation. 
{\em J. Theor. Biol.} {\bf 184}, 51-64.

  Gavrilets, S. \& Hastings, A. 1996. Founder 
effect speciation: a theoretical reassessment. {\em Am. Nat.} {\bf 147}, 
466-491. 

   Higgs, P.G. \& Derrida, B. 1992. Genetic distance and species
formation in evolving populations. {\em J. Mol. Evol.} {\bf 35}, 454-465.

  Hoelzer, G.A. \& Melnick, D.J. 1994. Patterns of speciation and
limits to phylogenetic resolution. {\em Trends in Ecology and Evolution}
{\bf 9}, 104-107.

  Johnson, T.C. et al. 1996. Late Pleistocene desiccation of 
Lake Victoria and rapid evolution of
cichlid fishes. {\em Science} {\bf 273}, 1091-1093.

  Kimura, M. \& Weiss, G.H. 1964. The stepping stone model of
population structure and the decrease of genetic correlation with distance.
{\em Genetics} {\bf 49}, 561-576.

  Li, W.-H. 1976. Distribution of nucleotide differences between
two randomly chosen cistrons in a subdivided population: the finite island
model. {\em Theor. Popul. Biol.} {\bf 10}, 303-308.
 
  Mallet, J. 1995. A species definition for the 
modern synthesis. {\em Trends Ecol. Evol.} {\bf 10}, 294-299.

  Manzo, F. \& Peliti, L. 1994. Geographic speciation in the
Derrida-Higgs model of species formation. {\em J. Phys. A: Math. Gen.} {\bf 27},
7079-7086.

  Maynard Smith J. \& E. Sz\"{a}thm\'{a}ry. 1995. {\em The Major Transitions
in Evolution}. Oxford: W.H. Freeman/Spektrum. 

  Mayr, E. 1942. {\em Systematics and the origin of
species}. New York: Columbia University Press.

  Mayr, E. 1963. {\em Animal Species and Evolution}, 
Cambridge: Harvard University Press.

  Mayr, E. 1982. {\em The Growth of Biological Thought: 
Diversity, Evolution, and Inheritance}. Cambridge: Belknap Press. 

  McCune, A.R. 1996. Biogeographic and stratigraphic evidence 
for rapid speciation in semionotid fishes. {\em Paleobiology} {\bf 22},
34-48.

  Muller, H.J. 1942. Isolating mechanisms, 
evolution and temperature. {\em Biol. Symp.} {\bf 6}, 71-125.

  Nei, M., Maruyma, I. \& Wu, C-I. 1983. Models of evolution of
reproductive isolation. {\em Genetics} {\bf 103}, 557-579. 

Noest, A.J. 1997. Instability of the sexual continuum. {\em Proc. R. Soc.
Lond. B} {\bf 264}, 1389-1393.

  Orr, H.A. 1995. The population genetics of 
speciation: the evolution of hybrid incompatibilities. {\em Genetics} 
{\bf 139}, 1803-1813.

  Orr, H.A. \& Orr, L.H. 1996. Waiting for speciation: the effect
of population subdivision on the waiting time to speciation. {\em Evolution}
{\bf 50}, 1742-1749.

  Reidys, C.M., Stadler, P.F. \& Schuster, P. 1997. Generic
properties of combinatory maps: neutral networks of RNA secondary
structures. {\em Bull. Math. Biol.} {\bf 59}, 339-397.

   Reinthal, P.N., and A. Meyer. 1997. Molecular phylogenetic tests of
speciation models in Lake Malawi cichlid fishes. Pp 375-390 {\em in} 
Givnish, T.J. \& K.J. Sytsma (eds). {\em Molecular evolution and 
adaptive radiation}. Cambridge: Cambridge University Press.

  Rosenzweig, M. 1995. {\em Species Diversity in Space and Time}.
Cambridge: Cambridge University Press.

  Schluter, D. 1996. Ecological causes of adaptive radiation.
{\em Amer. Natur.} {\bf 148}, S40-S64.

  Schluter, D. \& McPhail, J.D. 1992. Ecological displacement
and speciation in sticklebacks. {\em Amer. Natur.} {\bf 140}, 85-108.

  Singh, R.S. 1990. Patterns of species divergence and genetic
theories of speciation. Pp 231-265 {\em in} W\"{o}hrmann, K. \& S.K. Jain
(eds). {\em Population Biology. Ecological and Evolutionary Viewpoints}.
Berlin: Springer-Verlag.

  Slatkin, M. 1987. The average number of sites segregating DNA
sequences drawn from a subdivided population. {\em Theor. Popul. Biol.} {\em
32}, 42-49.

  Strobeck, C. 1987. Average number of nucleotide differences in a
sample from a single subpopulation: a test for population subdivision. 
{\em Genetics} {\bf 117}, 149-153.

  Vose, M.D. 1991. A linear algorithm for generating random numbers
with a given distribution. {\em IEEE Transactions on Software Engineering}
{\bf 17}, 972-974.

Wake, D.B. 1997. Incipient species formation in salamanders of the {\em
Ensatina} complex. {\em Proc. Natl. Acad. Sci. USA} {\bf 94}, 7761-7767.

  Wagner A., Wagner, G.P. \& Similion, P. 1994.
Epistasis can facilitate the evolution of reproductive isolation by 
peak shifts - a two-locus two-allele model. {\em Genetics} 138, 533-545.

  Watterson, G.A. 1975. On the number of segregating sites 
in genetic models without recombination. {\em Theor. Popul. Biol.} {\bf 7},
256-276.

  Whitlock, M.C.,  Phillips, P.C. , Moore, F.B.-G. \& Tonsor, S.J. 1995.
Multiple fitness peaks and epistasis. {\em Annu. Rev. Ecol. Syst.} {\bf 26},
601-629.

  Wright, S. 1932. The roles of mutation, inbreeding,
crossbreeding and selection in evolution. {\em Proc. Sixth
Int. Congr. Genet.} {\bf 1}, 356-366.

\newpage
\section{Legends}

Figure 1. Three types of dynamics. The population is subdivided into
two subpopulations. Of the three curves apparent
after generation 1000, the upper curve gives the AGD
between the subpopulations, whereas the two remaining curves give 
the AGD within each subpopulation. The total  
population size $N$, and $K$ value, are given on the graphs.
The arrows mark generation 1000 when restrictions on migration
($m=0.01$) we introduced.\\

Figure 2. Effects of population size and the strength of selection
on evolutionary divergence and speciation. The population is subdivided 
into two subpopulations. AGD between the subpopulations are shown 
(30 runs for each parameter configuration). 
({\em a}) With $K=10$ and $N=200$, subpopulations steadily diverged 
immediately after generation 1000 in all runs. 
({\em b}) However, with $N=400$ the time until 
the initiation of divergence varied and divergence was not observed 
in 13 runs. Comparing the slopes of the curves representing AGD
between subpopulations in ({\em a}) and ({\em b}) shows that
increasing population size decreases the rate of divergence as well.  
({\em c}) With $K=30$ and $N=200$ speciation was observed in only 2 runs. 
({\em d}) With $N=400$, speciation took place in 18 runs.
Other parameters are the same as in Fig.1.\\

Figure 3. Effects of spatial subdivision ($N=800, K=20$; number of
loci and the rates of recombination and mutation are the same as in Fig.1).
Figures ({\em a}) and ({\em b}) show AGD within and between subpopulations 
for a single run. Results for the 29 other runs are very similar. 
({\em a}) Population is subdivided into 2 equal subpopulations with migration 
rate $m=0.01$. No speciation is observed (cf. Fig.1a).
({\em b}) Population is subdivided into 4 equal subpopulations; island model 
with $m=0.01$. All AGD between subpopulations steadily increase after
generation 1000 indicating emergence of 4 species (cf. Fig.1b).
Figure ({\em c}) corresponds to a stepping-stone model with 8 local subpopulations with migration rate $m/2=0.03$ between neighboring subpopulations. 
AGD between neighboring subpopulations for a single run resulting in the emergence of two species are shown. 
All distances remain below $K$ (no reproductive isolation between 
neighboring subpopulations) except between subpopulations 4 and 5 which 
starts increasing around generation 2300 (cf. Fig.1c) indicating the split
into two species with 4 subpopulations each. \\

Figure 4. Splitting patterns in a stepping-stone
system with 8 local subpopulations. All parameters are the same as in
Fig.3c. ({\em a}) The observed distribution of the position of the first
break using data from 70 runs (out of 92 total runs) that resulted in
speciation. The abscissa gives the pair of subpopulations between which the 
split occurred. ({\em b}) Shown are the AGD within  the subpopulations 
(the bars above the 
integers from 1 to 8) and between the subpopulations (the bars between the 
integers) calculated
over generations 1500 through 3000 in 22 runs that did not result in
speciation. ({\em c}) The AGD between the first subpopulation and other
7 subpopulations using the same data as in ({\em b}).

\end{document}